\documentclass[10pt,conference]{IEEEtran}
\IEEEoverridecommandlockouts

\usepackage{cite}
\usepackage{amsmath,amssymb,amsfonts}
\usepackage{algorithmic}
\usepackage{graphicx}
\usepackage{textcomp}
\usepackage{xcolor}
\def\BibTeX{{\rm B\kern-.05em{\sc i\kern-.025em b}\kern-.08em
    T\kern-.1667em\lower.7ex\hbox{E}\kern-.125emX}}

\usepackage{amsmath,amsfonts,amsthm}
\usepackage{algorithmic}
\usepackage{graphicx}
\usepackage{textcomp}
\usepackage{xcolor}
\usepackage{verbatim}
\usepackage{etoolbox}
\usepackage{booktabs}
\usepackage{multirow}
\usepackage{braket}
\usepackage{quantikz}
\usepackage{tcolorbox}
\usepackage{verbatim}
\usepackage{etoolbox}
\usepackage{stmaryrd}
\usepackage{subcaption}
\usepackage[ruled,vlined]{algorithm2e}
\usepackage{stfloats} 
\usepackage{adjustbox}
\usepackage[normalem]{ulem}
\usepackage{cite}
\usepackage{amsmath,amssymb,amsfonts}
\usepackage{algorithmic}
\usepackage{graphicx}
\usepackage{textcomp}
\usepackage{xcolor}
\usepackage{capt-of}  
\usepackage{cuted}
\usepackage{multicol}
\usepackage{hyperref}
\usepackage{cleveref}

\begin{document}

\title{Design and Benchmarking\\of a Quantum Photonic Chip}

\author{
\IEEEauthorblockN{
Gabriele De Angelis\textsuperscript{5},
Nicol\`o Leone\textsuperscript{1,6},
Alessandro Luongo\textsuperscript{2,3,*},
Alberto Montanaro\textsuperscript{5},
Matteo Sanna\textsuperscript{1},\\
Roberto Siagri\textsuperscript{1,*},
Vito Sorianello\textsuperscript{5},
Luigi Tallone\textsuperscript{5},
Fabrizio Tamburini\textsuperscript{1}, and
Marco Venere\textsuperscript{4,3,6}
}
\IEEEauthorblockA{\textsuperscript{1}\textit{Rotonium S.r.l.}, Italy}
\IEEEauthorblockA{\textsuperscript{2}\textit{Centre for Quantum Technologies, National University of Singapore}, Singapore}
\IEEEauthorblockA{\textsuperscript{3}\textit{Inveriant Pte. Ltd.}, Singapore}
\IEEEauthorblockA{\textsuperscript{4}\textit{Department of Electronics, Information and Bioengineering, Politecnico di Milano}, Italy}
\IEEEauthorblockA{\textsuperscript{5}\textit{Photonic Networks e Technologies National Laboratory, CNIT}, Italy}
\IEEEauthorblockA{\textsuperscript{6}Equal contribution.}
\IEEEauthorblockA{\textsuperscript{*}Corresponding authors: \texttt{ale@inveriant.com}, \texttt{roberto.siagri@rotonium.com}}
}

\maketitle

\begin{abstract}
    We present the design and benchmarking of a quantum photonic processor, capable of encoding a quantum system in the degrees of freedom of single photons, based on standard CMOS-compatible manufacturing processes, and working at room temperature. We benchmark it against machine learning tasks, evaluating three quantum-classical architectures of increasing complexity. Our experimental results and simulations show that our chip achieves higher accuracy than classical networks of comparable size in multiple use cases. Compared to a superconducting quantum processor, it exhibits superior noise tolerance. These findings validate our steps toward a feasible, scalable route to efficient quantum applications and chip design.
    \end{abstract}

\begin{IEEEkeywords}
Quantum Computing, Photonic Chip, QPU benchmarking, Quantum Machine Learning.
\end{IEEEkeywords}

\section{Introduction}

Photonic quantum computing is emerging as one of the most promising paradigms for scalable, room-temperature quantum information processing. Yet, realizing large-scale photonic processors faces multiple challenges. In particular, we remark the difficulty of engineering effective photon–photon interactions for entangling gates, and the need for 
scalable architectures that can host and manipulate high-dimensional quantum states~\cite{bartolucci2023fusion,madsen2022quantum}. Finally, benchmarking pipelines should assess the performance of the devices, in terms of low-level error rates and high-level functional accuracy~\cite{proctor2025benchmarking}.

In this work, we present the design and benchmarking of $\mathsf{RP000}$, a three-qubit photonic quantum processor operating in the telecom C-band. We validate the processor at several levels of the quantum hardware–software stack: from the calibration of on-chip unitaries, to the execution of a multilayer \textit{Ansatz}, and finally to its use in supervised machine-learning tasks. We discuss how this chip implements a graded algebraic structure in which a single photon encodes multiple logical degrees of freedom, enabling compact multi-qubit operations and mode-preserving linear-optical transformations. 
The chip leverages a minimal $(\mathbb{Z}_2)^3$-graded Lie algebra to model and implement quantum photonic operations. The proposed graded-symmetry framework also provides a natural language for describing the coupling between optical modes, while ensuring algebraic consistency across the different parity sectors. For benchmarking, we devise a hardware/software co-design for quantum machine learning (ML) tasks~\cite{rotosoftware}, that compares theoretical predictions, numerical simulations, and experimental results. We also compare our performance with classical models and with an equivalent implementation on a superconducting quantum device. Taken together, our results show that the chip faithfully implements the target circuit, and achieves comparable or superior performance compared to other hardware.


We remark that the goal of this work is to identify design methodologies for photonic processors and production pipelines, rather than focusing on architectural scalability. Furthermore, this work also serves as a test vehicle for assessing the integration of design, fabrication, calibration, and benchmarking workflows. The primary barriers to reaching larger systems are engineering challenges, including detector and single photon source integration, control electronics density, and active stabilization of larger reconfigurable circuits, rather than fundamental physical limitations of the photonic platform. We will discuss scalability more in detail in future work.
\section{Preliminaries \& Related Works}\label{sec:prel-rel}
As this work spans multiple disciplines, we invite the interested reader to rely on~\cite{nielsen2010quantum,saleh2019, gerry2023introductory, wang2020} for background on quantum computation and optics. Moreover, some insights require background knowledge in abstract algebra -- an unfamiliar reader may find~\cite{majid2000foundations,klimyk2012quantum} useful for Hopf algebras, quantum groups, and related representation theory. 

\paragraph{Photonic Devices}\label{ssec:photonics}
integrated photonics~\cite{saleh2019} enables the manipulation of the light on compact, scalable, and power-efficient lithographic chips, called photonic integrated circuits (PICs). The fundamental building block of a PIC is the single-mode waveguide: light is confined in the waveguide core by total internal reflection arising from the refractive-index contrast between core and cladding material. Light is routed and processed on-chip by a small set of integrated components; the most common are: multimode interferometers (compact on-chip beam splitters/combiners), directional couplers (wavelength- and geometry-dependent couplers), waveguide crossings, phase shifters (implemented thermally or electro-optically), and Mach–Zehnder interferometers. These elements are combined with polarization transducers, edge or grating couplers for fiber I/O, and detectors to realize complete photonic circuits.
Integrated photonics is particularly well suited for quantum applications: the same lithographic primitives developed for classical telecom devices can be used, with appropriate control and readout, to manipulate single photons. This allows for reproducible, low-loss implementations of state preparation, controlled interferometry, and projective measurements in a manufacturable platform \cite{baldazzi2025,wang2018,adcock2019,llewellyn2020,adcock2020,baldazzi2024}. A broader overview on this topic is given by~\cite{wang2020}.

\paragraph{Graded Symmetry and Lie Algebra in Photonics}

let $\mathcal{M}$ denote the single-photon mode space. The bosonic Fock space built on $\mathcal{M}$ decomposes as
\(
\mathcal{F}(\mathcal{M})=\bigoplus_{N=0}^{\infty}\mathrm{Sym}^N(\mathcal{M}),
\)
where $\mathrm{Sym}^N(\mathcal{M})$ is the symmetric subspace of $\mathcal{M}^{\otimes N}$. The one-photon sector is canonically isomorphic to the mode space itself, i.e. $\mathcal{H}^{(1)} \cong \mathcal{M}$. For any number $n$ of qubits, or more generally for an $n$-level qudit, the operator algebra on the corresponding Hilbert space can be organized as an algebra graded by $\mathbb{Z}_2$ or by the minimal nontrivial grading group $G = \mathbb{Z}_2 \times \mathbb{Z}_2$. In this latter case, the graded Lie algebra 
$\mathfrak{g} = \bigoplus_{(a,b)\in\{0,1\}^2} \mathfrak{g}_{(a,b)}$
is partitioned into four homogeneous sectors and is endowed with the graded bracket $[X,Y] \;=\; XY - (-1)^{\, aa' + bb'}\, YX$, with $X \in \mathfrak{g}_{(a,b)}$ and $Y \in \mathfrak{g}_{(a',b')}$.
This bracket unifies commuting (even) and anticommuting (odd) behavior inside a 
single algebra and obeys the graded Jacobi identity. The $\mathbb{Z}_2 \times \mathbb{Z}_2$-graded Lie algebra is used as a convenient symmetry language to classify and describe degrees of freedom such as orbital angular momentum and spin-angular momentum interactions in a spin–orbit–coupled waveguide~\cite{Toppan2021,Tamburini2025GradedParaparticle}.

\begin{figure*}[htbp]
    \centering
    \includegraphics[width=\linewidth]{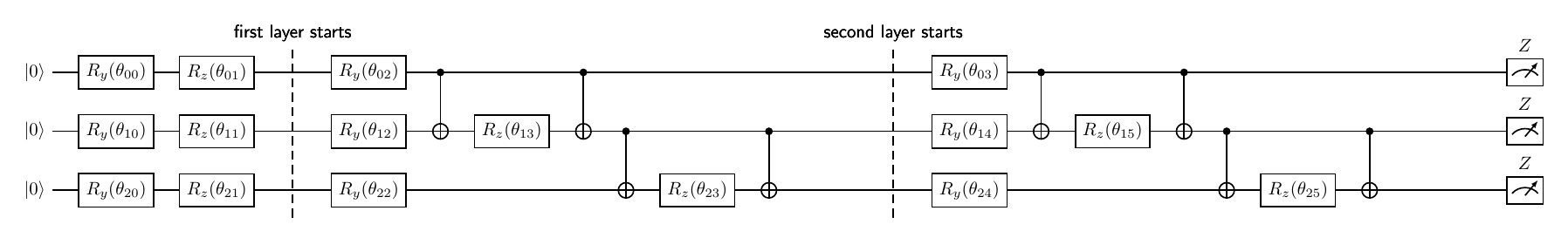}
    \caption{The \textit{Ansatz} implemented by $\mathsf{RP000}$ which we consider in our experiments. It consists of a layered architecture, where a series of $R_y$ and $R_z$ rotations alternate and $\texttt{CNOT}$ gates generate entanglement. }
    \label{fig:ansatz}
\end{figure*}

\paragraph{Quantum Benchmarking}
quantum hardware performance can be assessed at multiple layers of the software stack~\cite{proctor2025benchmarking,lorenz2025systematic}. At the component level, benchmarks characterize qubits and gates through coherence times ($T_1$, $T_2$), gate fidelities, tomography, randomized benchmarking, and cross-talk analysis~\cite{eisert2020quantum,knill2008randomized, magesan2012characterizing,flammia2021averaged,hockings2025scalable}. At the system level, benchmarks evaluate entire circuits using measures such as quantum volume~\cite{cross2019validating}, error-per-layered-gate~\cite{mckay2023benchmarking}, GHZ-state fidelity~\cite{toth2005entanglement}, and execution throughput. The software and HPC level targets the compiler and runtime stack, measuring transpilation efficiency, cross-device performance~\cite{hinsche2024efficient}, and overheads introduced by error-correction, or computational throughput (e.g., LINPACK~\cite{dongarra2003linpack}, RACBEM~\cite{dong2021random}). Most relevant to our case is the application level, which considers end-to-end performance, including data pre-processing, algorithmic computation, low-level error correction, and data post-processing. This contains complete software applications that rely on quantum hardware or simulators, e.g., QFT circuits~\cite{yin2021quantum} and simulation of physical systems~\cite{brown2010using}. Useful metrics include~\cite{martiel2021benchmarking,mesman2024quas,lubinski2023application,pollard2025benchqc,finvzgar2022quark},  differing in the scope, number of qubits, and depth of the circuits. Refer to~\cite{notton2025establishing,madsen2022quantum} for benchmarking of photonic devices.

\section{Chip Design}\label{sec:chipdesign}
The chip is a three-qubit quantum processor designed for operation in the telecom C-band. The chip is realized on a 220-nm silicon-on-insulator platform using standard CMOS-compatible manufacturing processes. This processor operates entirely at room temperature, encodes three qubits in the degrees of freedom of single photons, and manipulates them. This approach offers two major advantages: it eliminates the need for photon-photon interactions, and photon loss manifests as a reduction in detection rate rather than as logical errors that distort the measured probability distribution. Hence, photon loss increases the required acquisition time rather than corrupting the measurement statistics. The processor is fabricated on a commercial silicon-photonics multi-project wafer service and packaged to ensure stable operation. The on-chip components include Mach–Zehnder interferometers, which serve as the primary building blocks for implementing parametric rotations, and thermo-optic phase shifters, metallic heaters integrated above the waveguides, providing voltage-controlled phase tuning over the full $0-2\pi$ range. Light is coupled in and out of the chip through butt-coupled edge couplers interfaced to $127 \mu$m pitch polarization-maintaining fiber arrays in TE polarization. The chip package further integrates a thermoelectric cooler regulated by a proportional-integral-derivative (PID) feedback loop, ensuring thermal stability throughout extended measurement campaigns.
The processor is controlled electronically via a multichannel current driver that actuates the on-chip heaters. Single photons are generated by a heralded C-band source: the heralding photon is routed directly to the detection system, while the heralded photon is injected into the PIC and manipulated by the reconfigurable circuit. At the output, the processed photons are collected and detected by InGaAs single-photon avalanche detectors, with coincidence events time-resolved by a time-tagger module for high-precision photon-correlation measurements. Specifically, the probability of observing the state $|xyz\rangle$ is computed as the coincidence counts recorded within the chosen coincidence window for that state, $N_{xyz}$, normalized by the total number of coincidence counts $\mathbf{P}(|xyz\rangle)=\frac{\eta_{xyz}N_{xyz}}{\sum_{x,y,z}\eta_{xyz}N_{xyz}}$
where $\eta_{xyz}$ are the efficiencies in the photon-collection. 

\paragraph{Logical Circuit}
we show in \Cref{fig:ansatz} the logical circuit implemented by the chip. This parametric circuit, in jargon \textit{Ansatz} or parametric quantum circuit (PQC),
consists of multiple layers that embed parametric $R_y$ and $R_z$ rotations, combined with entanglement via $\texttt{CNOT}$ gates.
Specifically, a preliminary layer of $R_y$ and $R_z$ gates is applied to three logical qubits initialized as $\ket{0}$. Then, two identical layers follow, which combine parametric rotations and entanglement.
We devise this circuit for integration in a quantum ML pipeline, to perform data classification. Indeed, parametric rotations allow for both input encoding and model parameter training.

\paragraph{Calibration}
to ensure accurate and precise quantum operations, we perform a detailed calibration on the chip. Each phase shifter is tuned to cancel the native parasitic phase errors introduced by fabrication imperfections, since uncorrected errors would spoil the processor output. Calibration proceeds by sweeping the control power applied to each phase shifter, firstly individually, and then in combinations, while recording the corresponding coincidence counts. We then extract the phase-power relation, by interpolating each measured coincidence count, and choose the heater setpoint that produces the target phase (typically the fringe center or the phase required by the target unitary). We repeat this sweep-fit-set cycle iteratively until the entire PIC is under stable control. To validate the calibration procedure, a set of target unitaries designed to route the input photon deterministically into each of the eight computational basis states is implemented and tested. For all configurations, the measured probability of the target output state exceeds 98\%, with an average of $99.04\pm0.01\%$, with typical residual probabilities on non-target channels consistently below 1\%. Furthermore, repeated measurements over a continuous 64-hour campaign show no performance degradation, with target-state probability fluctuations confined within $\pm 2\%$ at $3\sigma$. No recalibration was required throughout the entire acquisition.

\paragraph{Chip as Graded Paraparticle}
this photonic chip implements three logical qubits by mapping momentum labels to grades, so each single photon carries $3$ qubits spanning the four algebraic graded sectors. We extend previous work~\cite{Tamburini2025GradedParaparticle} to show that the same formalism presented in Section~\ref{sec:prel-rel} can be used to describe computation encoded in photon's momentum only. When $n$ is a power of two, $n=2^m$, there is a canonical qubit tensor: $\mathcal{H}_n \ \cong\ (\mathbb{C}^2)^{\otimes m}$.
If $n=2^m$, after fixing a binary labeling, there is a (basis–dependent) identification $\mathcal H_n\cong(\mathbb C^2)^{\otimes m}$. If $n=\prod_{r=1}^R n_r$ with pairwise coprime $n_r$, the Chinese Remainder Theorem (CRT) gives $\mathbb Z_n\cong\prod_r \mathbb Z_{n_r}$ and $\mathbb C^n\cong\bigotimes_r \mathbb C^{n_r}$. On $\mathcal H_n$ the Heisenberg–Weyl pair is 
$Z_n\ket{a}=\omega^{a}\ket{a}$, $X_n\ket{a}=\ket{a{+}1\bmod n}$, $\omega=e^{2\pi i/n}$, and under CRT one has $X_n\mapsto\bigotimes_r X_{n_r}$ and
$Z_n\mapsto\prod_r Z_{n_r}^{\,M_r}$ with $N_r:=n/n_r$ and $M_r N_r\equiv 1\ (\mathrm{mod}\ n_r)$.
If $n$ is not a power of two, a convenient binary reshaping proceeds one $\mathbb Z_2$ at a time, $\ket{j}\mapsto\ket{j\bmod 2}\otimes\ket{\lfloor j/2\rfloor}$ and $\mathbb C^n \hookrightarrow \mathbb C^2\otimes\mathbb C^{\lceil n/2\rceil}$,
and iterating yields an isometric embedding into $m=\lceil\log_2 n\rceil$ qubits (with a partially
filled top block when $n$ is odd). In our momentum–only chip, two binary momentum parities label the $\mathbb Z_2\times\mathbb Z_2$ grades and a third binary mode is the intra-grade fiber. We believe this formalism can be useful for syndrome extraction using parity checks, as required by most quantum error correction codes. 

\section{Benchmarking Methodology}\label{sec:methodology}

To evaluate the performance of the chip, we design a testbed that (\textbf{i}) validates the physical chip by quantifying how closely its implemented operations match the theoretical specification, (\textbf{ii}) compares the chip with a superconducting architecture implementing the same circuit, and (\textbf{iii}) characterizes the performance of the circuit at the \emph{application level}. To this end, we devise a quantum ML pipeline for data classification, that embeds our quantum chip. Specifically, we study its learning dynamics in both a purely quantum classification task and a hybrid quantum–classical setting, to assess how far its practical power can be extended beyond standard quantum-only use cases -- and how it compares against purely classical baselines. As a byproduct, these experiments are also measuring \emph{how many} trainable parameters our quantum circuit needs to maximize accuracy, and what degree of accuracy is achieved by a classical neural network of comparable size. Indeed, despite the known limitations of variational circuits~\cite{cerezo2025does,nemkov2025barren,abbas2023quantum}, a lower number of parameters has the potential to imply a lower execution time for optimization. 

\subsection{Ansatz Training and Validation Process}\label{subsec:training}

\begin{figure}[t]
    \centering
    \includegraphics[width=\linewidth]{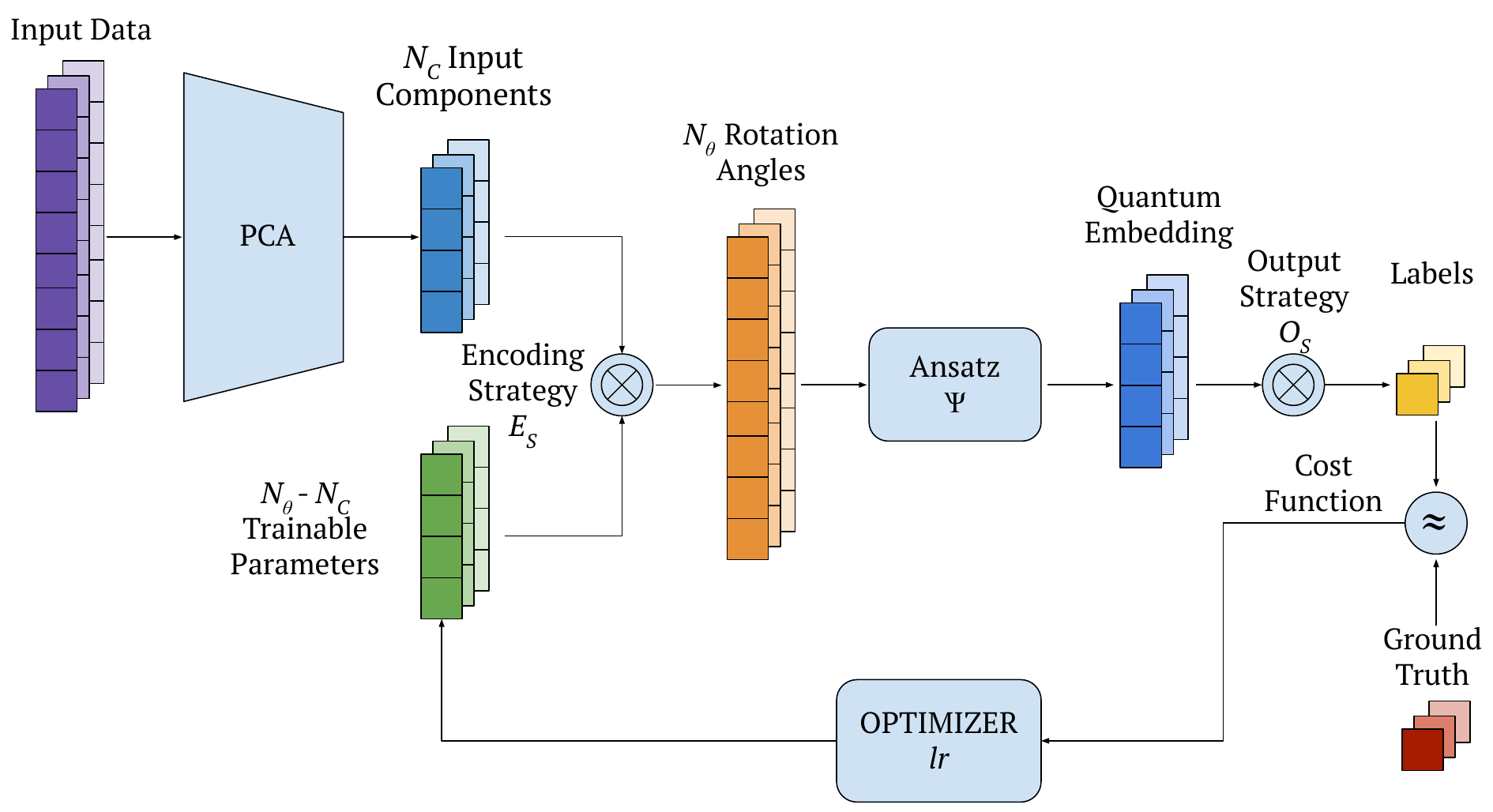}
    \caption{Description of the training pipeline for our quantum \textit{Ansatz}. First, data is pre-processed via PCA; then, the extracted $N_C$ components are combined with trainable parameters via an encoding strategy $E_S$, to produce $N_\theta$ rotation angles that feed the \textit{Ansatz}. The expectation values of the 3 Pauli $Z$ measurement operators will determine our \textit{quantum embedding} of data. 
    After the labels are derived, we evaluate the considered angles and invoke the Adam optimizer.} 
    \label{fig:ansatz-training}
\end{figure}

As shown in \Cref{fig:ansatz-training}, to perform classification we envision a hardware-software co-design that classically pre-processes data, produces the rotation angles for the \textit{Ansatz}, simulates or executes the produced quantum circuit to retrieve a \textit{quantum embedding}, and evaluates its accuracy to further optimize parameters.

\paragraph{Classical Pre-Processing}
    given the batch of data to feed as input for training and inference, we reduce its dimensionality through principal component analysis (PCA)~\cite{gewers2021principal}. This is a technique for dimensionality reduction that, given some input data with a high number of features, analyzes their distribution and performs a linear transformation. This new coordinate system will capture the largest variance across the data, so as to reduce the number of features while preserving, at least partially, data distinguishability. This stage is necessary as the number of angles for data encoding in the logical circuit is limited. In this work, we refer to the number of components that we extract with PCA as $N_C$.
    
\paragraph{Encoding} this is a software module that generates $N_\theta$ rotation angles for the \textit{Ansatz}, starting from the $N_C$ components. Indeed, as described in \Cref{sec:chipdesign}, rotation angles can be employed for both input encoding and trainable parameters. Since different alternation strategies of input and trainable parameters in the circuit will generally lead to different learning capabilities, as we will show in \Cref{sec:results}, we devise and explore multiple encoding strategies: \texttt{$R_y$\_input} (\texttt{$R_z$\_input}), encoding input in the $R_y$ ($R_z$) gates, respectively; \texttt{$R_y$\_$R_z$\_input}, which employs both $R_y$ and $R_z$ gates for input encoding; \texttt{input\_first} (\texttt{param\_first}), which schedules all input angles (parameter angles) before parameter angles (input angles), respectively. Given $N_C$ and $N_\theta$, the strategy may employ data re-uploading to fit the number of angles and improve learning~\cite{perez2020data}. 
    
\paragraph{Execution \& Output} given the rotation angles, we generate the final quantum circuit and perform either simulation or execution on real quantum hardware. During simulation, we derive the theoretical expectation values of the Pauli $Z$ measurements, while we estimate them via sampling in the case of execution on hardware. We remark that executing on hardware requires statevector tomography, with a number of measurements proportional to the desired accuracy. Based on Theorem 4.5 from Reference~\cite{kerenidis2019quantum}, given an estimation error $\varepsilon$ and $2^n$ observables to measure, the number of required measurements is $\mathcal{\widetilde{O}}\left(\frac{n}{\varepsilon^2}\right)$. At this point, we derive a vector of three expectation values, one per qubit, which we refer to as the \textit{quantum embedding} that represents our input data. For a binary classification task, we may use the hyperparameter $O_S$ to select a given qubit and derive a classification label via a sigmoid function applied to the expectation value for that qubit. For hybrid models, we feed our \textit{quantum embedding} to a classical neural network, and derive the classification labels from $N_O$ classical output neurons.

\paragraph{Optimization \& hyperparameter tuning} given the output labels of the model, and the ground truth provided by the dataset, we optimize the  trainable parameters of the circuit to minimize a cross-entropy loss function, employing an Adam optimizer whose learning rate $lr$ is a further hyperparameter of the model. Furthermore, we perform hyperparameter tuning to identify the best configuration of hyperparameters for our considered model, as shown in \Cref{tab:hyperparameters}. 
Our training pipeline performs an $80\%-20\%$ training-test split, and then employs a $k$-fold cross-validation approach over the training set, to provide a more robust validation accuracy for hyperparameter tuning. The final test analysis, achieved with the best hyperparameter combination according to validation accuracy, is performed with a training over the full training set. We remark that this pipeline closely resembles how literature measures performance for quantum ML~\cite{bowles2024better}. 

\begin{table}[t]
    \centering
    \renewcommand{\arraystretch}{1.2}
    \caption{
    Description of the hyperparameters considered in our experiments and subject to tuning. \textbf{(a)} Hyperparameters for \textit{Ansatz} training, and related pre-, and post-processing. \textbf{(b)} Hyperparameters related to classical layers. We also place these hyperparameters in \Cref{fig:ansatz-training,fig:models}, where suitable.
    }
    \label{tab:hyperparameters}
    \begin{tabular*}{\linewidth}{@{\extracolsep{\fill}} lp{0.89\linewidth}}
        \toprule\toprule
        \textbf{Name} & \textbf{Description} \\
        \midrule
        $N_C$ & Number of components used to represent inputs. \\
        $E_S$ & Encoding strategy. Possible values: \texttt{$R_y$\_input}, \texttt{$R_z$\_input}, \texttt{$R_y$\_$R_z$\_input}, \texttt{input\_first}, and \texttt{param\_first}.\\
        $O_S$ & Output qubit. \\
        $lr$ & Learning rate. \\
        \midrule
        $H_S$ & Number of neurons in the hidden layer of in $\mathsf{Serial}_{QC}$ \\
        $H_{P}^{1}$ & Number of neurons in the first hidden layer in $\mathsf{Parallel}_{QC}$.\\
        $H_{P}^{2}$ & Number of neurons in the second hidden layer in $\mathsf{Parallel}_{QC}$.\\
        $H_C$ &  Number of neurons in the classical module that replaces the \textit{Ansatz} in classical counterparts.\\ 
        \bottomrule
    \end{tabular*}
\end{table}

\subsection{Machine Learning Models}

\begin{figure*}[t]%
  \centering
  \begin{minipage}[b]{0.48\textwidth}
    \centering
    \includegraphics[width=\linewidth]{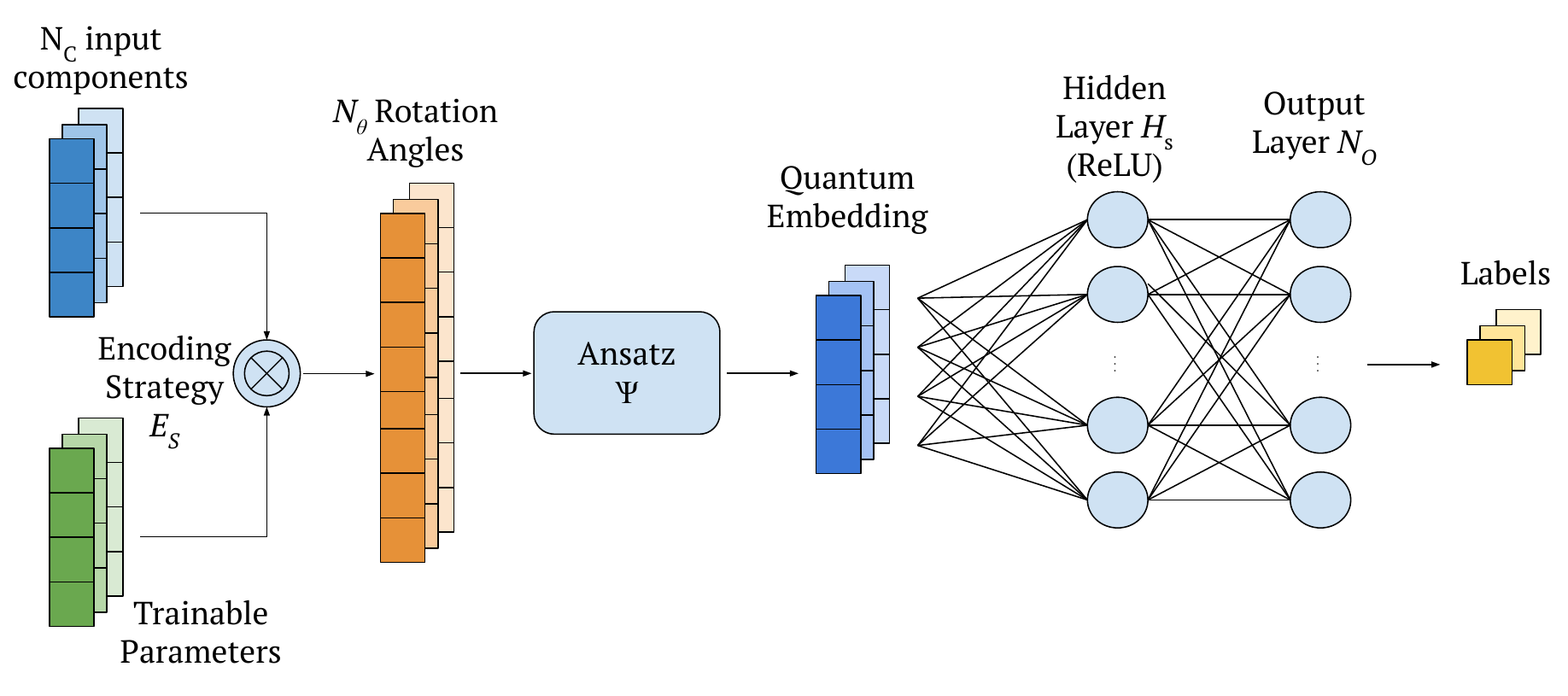}
    \subcaption{$\mathsf{Serial}_{QC}$: first, the \textit{Ansatz} produces a quantum embedding. Then, a classical network performs further processing to derive the output.}
    \label{subfig:model2-3}
  \end{minipage}\hfill
  \begin{minipage}[b]{0.48\textwidth}
    \centering
    \includegraphics[width=\linewidth]{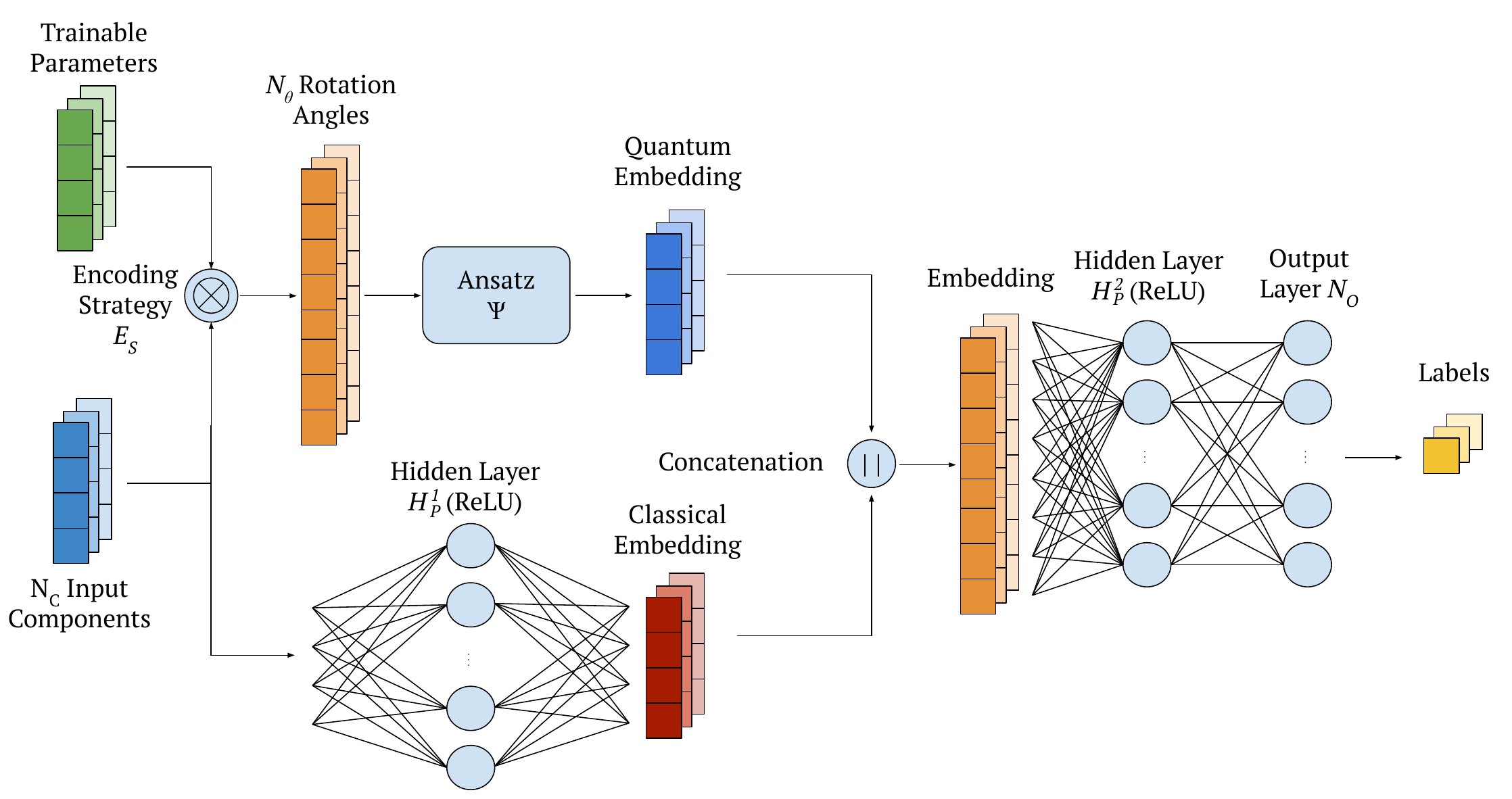}
    \subcaption{$\mathsf{Parallel}_{QC}$: a quantum \textit{Ansatz} and a classical network independently extract features from data; finally, a classical network is applied.}
    \label{subfig:model4}
  \end{minipage}
  \caption{Description of the \textit{hybrid} models considered by our experimental analysis, integrating a quantum \textit{Ansatz} into a more extended neural network. The quantum embedding is fed as input to a classical neural network, to provide higher capabilities in feature extraction and class separation in the embedding space. }
  \label{fig:models}
\end{figure*}


\paragraph{$\mathsf{QuantumNN}$}
this is a plain quantum model, based on the circuit of \Cref{fig:ansatz}, trained, validated, and tested as described in \Cref{subsec:training}. We employ it for binary classification tasks, by selecting a specific output qubit with $O_S$, and applying a sigmoid function on its corresponding expectation value. We then employ a binary cross-entropy loss function.

\paragraph{$\mathsf{Serial}_{QC}$}
to improve the learning capabilities of  $\mathsf{QuantumNN}$, we design a hybrid quantum-classical neural network, as shown in
\Cref{subfig:model2-3},  consisting of a series of a quantum and a classical layer. Specifically, the quantum embedding produced by the \textit{Ansatz} is fed as input to the subsequent fully connected layers, where the hidden layer has $H_S$ neurons. We first apply this scheme to a binary classification task, by setting the output layer dimension $N_O = 1$, using a sigmoid non-linearity, and the binary cross-entropy as the loss function.  Then, we extend this model to a multi-class classification problem, setting $N_O = M$, and feeding raw logits to the cross-entropy loss function. Our intuition is that, by feeding quantum embeddings as input to a classical neural network, the hybrid model can provide higher capabilities in feature extraction and class separation in the embedding space. 

\paragraph{$\mathsf{Parallel}_{QC}$}
we further augment the learning capabilities of $\mathsf{Serial}_{QC}$ by devising a parallel-series architecture, as shown in \Cref{subfig:model4}. Indeed, in the first stage, the $N_C$ input components are fed as input to both a quantum and a classical component, where the latter has a hidden layer of size $H_{P}^{1}$.  Then, the output embeddings of these components are concatenated and fed as input to a further classical component, with a hidden layer of $H_{P}^{2}$ neurons, that provides the final output of the problem. We employ this architecture to allow both a classical model and a quantum model to learn features from data, then combine them and derive a correlation between these embeddings and the output probabilities via a classical network.

\paragraph{Classical models} to assess the learning capabilities of our architectures, we compare their test accuracy against classical models. These models implement the classical counterpart of $\mathsf{QuantumNN}$, i.e., a classical neural network for binary classification, receiving as input $N_C$ components, computing a hidden layer of $H_C$ neurons with a \texttt{ReLU} non-linearity, and providing an output layer whose dimensionality $N_O = 1$ in case of binary classification, or $N_O = M$ in case of a multi-class experiment, where $M$ is the number of classes. The former case employs a binary cross-entropy loss function, while the latter case provides raw logits as output, which are fed to a cross-entropy loss function. This model is the software baseline we derive for comparison, and shall have a value $H_C$ such that its number of parameters matches, up to the minimum possible approximations, the number of parameters of the quantum solution.

\section{Experimental Results}\label{sec:results}
We run our numerical simulations on an AMD EPYC 7763 with 50 virtual cores and 60 GB of RAM, employing PyTorch 2.6.0, Pennylane 0.40.0, and Optuna 4.2.1. For hyperparameter tuning, we searched a space of up to 80 configurations. The experiments on the photonic architecture were executed on the chip, which is compared against the (noiseless) numerical simulation, and a superconducting chip. We consider a classification task, targeting both binary and multi-class classification on STATLOG~\cite{STATLOG}, a Landsat Satellite dataset, and YEAST~\cite{nakai1991expert}, a dataset for the localization of proteins.  
For the binary classification task, we select two well-separated and easily distinguishable classes, that is, classes 1 and 7 for STATLOG, and classes \texttt{CYT} and \texttt{NUC} for YEAST. 
For multi-class, we consider the whole dataset.



\begin{table*}[t]
    \centering
    \caption{
    Comparison of the accuracy achieved by the different models against their classical counterparts, i.e., replacing their quantum \textit{Ansatz} with a classical fully connected network of comparable size. The number of classical parameters only considers the classical network module replacing the \textit{Ansatz}. We provide the best hyperparameter configuration derived by tuning, their corresponding test accuracy (noiseless simulation, and $\mathsf{RP000}$ for \texttt{QuantumNN}), and the corresponding classical test accuracy.} 
    \label{tab:accuracies}
    \resizebox{\linewidth}{!}{
    \begin{tabular}{c|c|c|ccccccc|cc|cc|cc}
    \toprule \toprule
    \multirow{2}{*}{\textbf{Dataset}}&\multirow{2}{*}{\textbf{Task}}&\multirow{2}{*}{\textbf{Model}}&\multicolumn{7}{c|}{\textbf{Best Configuration}}&\multicolumn{2}{c|}{\textbf{Num. Parameters}}&\multicolumn{2}{c|}{\textbf{Quantum Accuracy}}&\multicolumn{2}{c}{\textbf{Classical Accuracy}}\\
    & & & $N_C$ & $E_S$ & $O_S$ & $lr$ & $H_S$ & $H_P^1$ & $H_P^2$ & Ansatz & Classical & $\mathsf{RP000}$ & Sim. & Avg. & Max. \\
    \midrule
    \multirow{4}{*}{STATLOG} & \multirow{2}{*}{binary} & \texttt{QuantumNN} & 8 & \texttt{$R_z$\_input} & 2 & 0.001 & - & - & - & 8 & 11 & 97.74\% & 99.35\% & 80.16\% & 99.57\%\\
    & & Serial$_{QC}$ & 6 & \texttt{$R_z$\_input} & - & 0.01 & 256 & - & - & 10 & 13 & - & 99.67\% &     99.29\% & 99.79\%\\ 
    \cmidrule(lr){2-15}
    & \multirow{2}{*}{multi-class} & Serial$_{QC}$ & 6 & \texttt{$R_y$\_input} & - & 0.01 & 256 & - & - & 10 & 13 & - & 76.95\% & 31.31\% & 31.31\%\\
    & & Parallel$_{QC}$ & 12 & \texttt{$R_y$\_input} & - & 0.001 & - & 256 & 512 & 4 & 19 & - &  76.10\% & 75.69\% & 80.75\%\\
    \midrule
    \multirow{4}{*}{YEAST} & \multirow{2}{*}{binary} & \texttt{QuantumNN} & 8 & \texttt{$R_y$\_$R_z$\_input} & 2 & 0.005 & - & - & - & 8 & 11 & 63.12\% & 62.56\% & 51.86\% & 64.25\%\\
    & & Serial$_{QC}$ & 6 & \texttt{$R_y$\_input} & - & 0.01 & 128 & - & - & 10 & 13 & - & 61.59\% & 56.63\%  & 57.54\%\\ 
    \cmidrule(lr){2-15}
    & \multirow{2}{*}{multi-class} & Serial$_{QC}$ & 8 & \texttt{$R_y$\_input} & - & 0.01 & 128 & - & - & 8 & 15 & - & 42.09\% & 31.31\% & 31.31\%  \\
    & & Parallel$_{QC}$ & 6 & \texttt{$R_y$\_input} & - & 0.01 & - & 128 & 128 & 10 & 13 & - & 44.44\% & 43.77\% & 44.78\%\\    
    \midrule
    \bottomrule
    \end{tabular}
    }
\end{table*}

\paragraph{Binary classification with $\mathsf{QuantumNN}$}\label{subsec:quantumnn}

first, we aim to evaluate the performance of our $\mathsf{QuantumNN}$ model, i.e., the plain \textit{Ansatz} model, and compare the chip accuracy to its noiseless simulation, and its classical counterpart, by training, validating, and testing with similar conditions and with a similar number of parameters for binary classification, and after performing hyperparameter tuning.

As shown in \Cref{tab:accuracies},  $\mathsf{QuantumNN}$ achieves a noiseless simulated accuracy of 99.35\% on the STATLOG dataset, employing 8 trainable parameters. A neural network of comparable size, employing only 1 hidden layer ($H_C = 1$) and 11 trainable parameters overall, achieves a maximum test accuracy of 99.57\% and an average test accuracy of 80.16\% over 30 runs. When run on the chip, we achieve 97.74\% accuracy. 
Regarding YEAST, we achieve a simulated test accuracy of 62.56\% with 8 quantum trainable parameters, compared to classical 51.86\% (avg.) and 64.25\% (max.) achieved with 11 trainable parameters, and 63.12\% on the chip. 
We remark that the chip undergoes noise effects which may either deteriorate or improve test accuracy compared to noiseless simulations.
Our experimental analysis shows that a quantum model is comparable to and, in some cases, outperforms classical models of similar size. This is not surprising, and has already been observed in the literature~\cite{bowles2024better}.

\begin{figure}[t]

        \centering
        \includegraphics[width=0.8\linewidth]{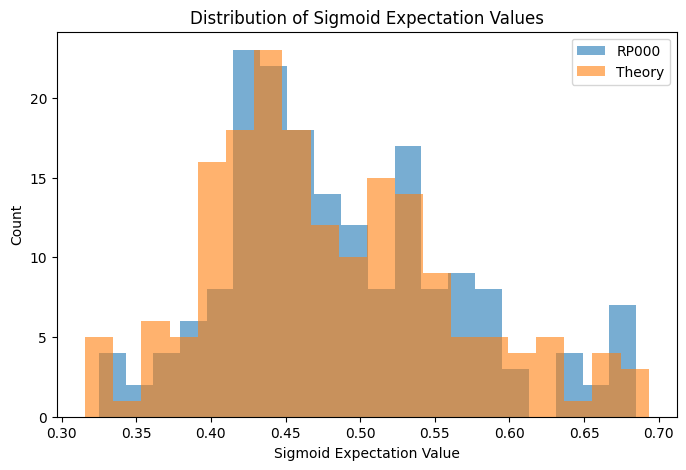}
    \caption{Comparison between the frequencies of the value of the sigmoids obtained by the numerical simulation (Theory) and the quantum chip ($\mathsf{RP000}$) for the test set of YEAST.}       \label{fig:expvalue_RP000_theory}
\end{figure}

We aim to estimate the effects of quantum noise and sampling accuracy when executing on the chip, compared to theoretical expectations and to a superconducting architecture. 
To do so, for a sample $i$ in the test set, we define $a_i=\sigma(Tr[O\rho_i])$, where $O$ is the observable associated with $O_S$, $\rho_i$ is the state of the quantum computer at the end of the circuit, and $\sigma$ is the sigmoid function. We measure the error on the sigmoid as $|a_i - \overline{a_i}|$, where $\overline{a}_i$ is the estimate obtained by sampling the quantum circuit. \Cref{fig:expvalue_RP000_theory} shows the frequencies of the different values of $a_i$ and $\overline{a}_i$, targeting the YEAST dataset on the chip. 
For a comparison against superconducting devices, we repeated our experiments on a 20-qubit superconducting architecture with transmon qubits connected via tunable couplers in a square lattice (${\sim}99.7$\% 1qb-gate and ${\sim}97.6$\% 2qb-gate median fidelity). We use the exact same circuit structure, rotation angles, and hyperparameter settings as in the photonic implementation. Importantly, since the superconducting device supports a larger qubit count, we selected the best qubits of the chip in terms of error rates maximizing 2 qubit fidelity. For the ML experiments, in order to assure consistency with the photonic chip, we kept the same number of measurement shots as in the photonic case, so that the comparison is made under matched sampling statistics. 
We further measured the errors using two distinct metrics: the mean absolute error (MAE) and the root mean square error (RMSE). Formally, $MAE = \frac{1}{N}\sum_i |a_i - \overline{a_i}|$ and $RMSE = \left(\frac{1}{N}\sum_i (a_i - \overline{a_i})^2\right)^{1/2}$.
For the STATLOG dataset, the chip achieves an MAE error of $0.0222$ and an RMSE $0.0275$, while the superconducting architecture achieves a MAE error of $0.0438$ and a RMSE $0.0527$; for the YEAST dataset, the chip has an MAE error of $0.0139$ and a RMSE $0.0169$, while superconducting  
has a MAE error of $0.0158$ and a RMSE $0.0201$. Both datasets, therefore, prove the superior performance of the chip. The test accuracy achieved by superconducting is 90.97\% on STATLOG and 63.68\% on YEAST.  


\paragraph{Hybrid quantum models}\label{sec:experiments-hybrid}

finally, we show in \Cref{tab:accuracies} the accuracy achieved by noiselessly simulating our hybrid models, compared to classical counterparts, on both datasets, for both tasks.
The $\mathsf{Serial}_{QC}$ model achieves a test accuracy of 99.67\% on the binary STATLOG problem using 10 quantum parameters, compared to the classical 99.29\% (avg.) and 99.79\% (max.) with 13 parameters. We remark that the presence of a hybrid architecture determines an improvement in accuracy, compared to the plain $\mathsf{QuantumNN}$ model, and outperforms the classical counterpart if we normalize the achieved accuracy with respect to the number of classical parameters. A similar trend occurs with the YEAST dataset, with the difference that in this case $\mathsf{Serial}_{QC}$ outperforms the classical counterpart, while is slightly less efficient than $\mathsf{QuantumNN}$.
Regarding the multi-class problem, $\mathsf{Serial}_{QC}$ outstandingly outperforms classical counterparts (76.95\% vs 31.31\% for STATLOG; 42.09\% vs 31.31\% for YEAST). Finally, our $\mathsf{Parallel}_{QC}$ architecture achieves 76.10\% accuracy on the multi-class STATLOG problem, compared to classical 75.69\% (avg.) and 80.75\% (max.), while reaches 44.44\% accuracy on the multi-class YEAST problem, compared to classical 43.77\% (avg.) and 44.78\% (max.). In this latter case, we remark that $\mathsf{Parallel}_{QC}$ outperforms $\mathsf{Serial}_{QC}$, thus showing the benefits of using both modules for feature extraction.
\section{Conclusions}\label{sec:concl}
We have validated a room-temperature, CMOS-compatible photonic quantum processor operating in the telecom C-band. By encoding three qubits into the degrees of freedom of single photons and organizing mode couplings within a $(\mathbb{Z}_2)^3$-graded Lie-algebraic framework, the chip realizes a compact, mode-preserving architecture. We benchmarked this architecture with a hardware-software co-design for QML experiments, considering both plain and hybrid models, and showed the achieved accuracy and error rates. Future work will extend this architecture to larger number of qubits, implementing and benchmarking explicit parity-check operations.


\section*{Acknowledgements}
Rotonium led the device design in collaboration with CNIT, while Inveriant designed the benchmarking and applications. FT developed the theoretical formalism for graded computation. AL and MV conducted the benchmarking experiments. NL, MS, VS, LT, AM contributed to experimental activities, coordinated by RS. We thank CQT for providing access to the superconducting hardware and Filippo Miatto for useful discussions. This work is partially supported by CQT Bridging Grant and the National Quantum Computing Hub (NQCH) 3.0 as part of the Quantum Engineering Programme (QEP) 3.0.

\bibliographystyle{IEEEtran}
\bibliography{bibliography}

\end{document}